\UseRawInputEncoding
\documentclass[aps,prc,groupedaddress,showkeys,showpacs,manuscript]{revtex4}
\usepackage{amssymb}
\usepackage{array}
\usepackage{graphicx}

\newcommand{\upcite}[1]{\textsuperscript{\textsuperscript{\cite{#1}}}}

\begin{document}

\title{Study of production of (anti)deuteron observed in Au+Au collisions at $\sqrt{s_{NN}}$=14.5, 62.4 and 200 GeV}

\author {Ying Yuan,$\, ^{1}$, $\, ^{2}$\footnote{E-mail address: yuany@gxtcmu.edu.cn}}
\address{
1) Mathematics and Physics Teaching and Research Section, College of Pharmacy, Guangxi University of Chinese Medicine, Nanning 530200, China\\
2) Guangxi Key Laboratory of Nuclear Physics and Nuclear Technology, Guangxi Normal University, Guilin 541004, China}


\begin{abstract}
Transverse momentum distributions of deuterons and anti-deuterons in Au+Au collisions at $\sqrt{s_{NN}}$=14.5, 62.4 and 200 GeV with different centralities are studied in the framework of the multi-source thermal model. Transverse momentum spectra are conformably and approximately described by the Tsallis distribution. The dependence of parameters (average transverse momenta, effective temperature and entropy index) on event centrality are obtained. It is found that the parameters $T$ increase and $q$ decrease with increase of the average number of particles involved in collisions, which reveals the transverse excitation degree increases with collision centrality.
\end{abstract}



\maketitle

{\section{Introduction}}

The study of strongly interacting matter at extreme temperatures and densities can be performed by colliding heavy-ion collisions at ultra-relativistic energies \upcite{Alt1,Sun2,Wang3,Li4,Liu5}. The production mechanism of hadrons and nuclei in ultra-relativistic heavy ion collisions deserves more investigation since it may give important message on the quantum chromodynamics (QCD) phase transition from quark-gluon plasma (QGP) to hadron gas (HG) \upcite{Arsen6,Li7}. The RHIC is scheduled to run at the energies which are around the critical energy of phase transition from hadronic matter to QGP \upcite{Lao8}. The theoretical study of nuclei and anti-nuclei has been undertaken for many years, for example the thermal model and coalescence model \upcite{Mrowc9,Mrowc10,Bazak11,Liu12,Liu13}. In particular, the study of transport phenomena is important for the understanding of many fundamental properties \upcite{Li14}. The spectra of transverse momentum of particles produced in high energy collisions are of high interest as they provide us an important information of the kinetic freeze-out state of the interacting system \upcite{Chen15}. At the stage of kinetic freeze-out, the effective temperature is not a real temperature, and it describes the sum of excitation degree of the interacting system and the effect of transverse flow \upcite{Waqas16}.

In this paper, using the Tsallis distribution \upcite{Tsallis17,Biro18,Cleymans19} in the multisource thermal model to simulate the transverse momentum distributions of (anti-)deuterons in Au+Au collisions at RHIC, we compare them with experiment data taken from the STAR Collaboration \upcite{Adam20}. The main purpose of this work is to extract the information on effective temperature, because it allows us to extract the kinetic freeze-out temperature.

\vspace{1\baselineskip}

{\section{The model and method}}

The model used in the present work is the multisource thermal model \upcite{Liu21,Liu22,Liu23}. In this model, many emission sources are formed in high-energy nucleus-nucleus collisions. The different distributions can describe the emission sources and particle spectra, such as the Tsallis distribution, the standard (Boltzmann, Fermi-Dirac and Bose-Einstein) distributions, the Tsallis+standard distributions \upcite{Buyukkilic24,Chen25,Conroy26,Pennini27,Teweldeberhan28,JConroy29}, the Erlang distribution \upcite{Liu21}, etc. The Tsallis distribution can be described by two or three standard distribution.

The experimental data of the transverse momentum spectrum of the particles are fitted by using the Tsallis distribution which can describe the temperature fluctuation in a few sources to give an average value. The Tsallis distribution has many function forms \upcite{Tsallis17,Biro18,Cleymans19,Buyukkilic24,Chen25,Conroy26,Pennini27,Teweldeberhan28,JConroy29,Zheng30,Zheng31}. In the rest frame of a considered source, we choose a simplified form of the joint probability density function of transverse momentum ($p_{\rm T}$) and rapidity ($y$) \upcite{Lao8},
\begin{equation}
f(p_{\rm T},y)\propto{\frac{d^{2}N}{dydp_{T}}}={\frac{gV}{{(2\pi)}^{2}}}{p_{T}}{\sqrt{p_{\rm T}^{2}+m_{\rm 0}^{2}}}{\cosh{y}}\times{{[1+{\frac{q-1}{T}}({\sqrt{p_{\rm T}^{2}+m_{\rm 0}^{2}}}{\cosh{y}}-{\mu})]}^{-{\frac{q}{q-1}}}}.
\label{eq:1}
\end{equation}
Here, $N$ is the particle number, $g$ is the degeneracy factor, $V$ is the volume of emission sources, $m_{0}$ is the rest mass of the studied particle, $T$ is the temperature which describes averagely a few sources (local equilibrium states), $q$ is the entropy index which describes the degree of non-equilibrium among different states, $\mu$ is the chemical potential which is related to $\sqrt{s_{NN}}$ \upcite{Andronic32}. In the RHIC energy region, the values of $\mu$ are shown in the Table I\upcite{Adamczyk33}. We can extract the values of $T$, $q$ and $V$ by reproducing the particle spectra, where $T$, $q$ are fitted independently for the studied particle, and $V$ is related to other parameters.

\begin{table}
\caption{Values of $\mu$ corresponding to the curves in Au+Au collisions at $\sqrt{s_{NN}}$=14.5 GeV, $\sqrt{s_{NN}}$=62.4 GeV and $\sqrt{s_{NN}}$=200 GeV for $0-10\%$, $10-20\%$, $20-40\%$, $40-60\%$ and $60-80\%$ centralities.} \label{Table1}
\begin{tabular}{ p{3cm}<{\centering} p{3cm}<{\centering} p{3cm}<{\centering} p{3cm}<{\centering}}
\hline \hline
      $\sqrt{s_{NN}}$ (GeV) & cross section & $\mu$ (MeV) \\
\hline
      14.5 & 0-10\% & $288.9\pm12.9$ \\
           & 10-20\% & $284.9\pm12.9$ \\
           & 20-40\% & $278.7\pm12.8$ \\
           & 40-60\% & $256.0\pm12.4$ \\
           & 60-80\% & $227.3\pm1.08$ \\
\hline
      62.4 & 0-10\% & $66.1\pm5.3$ \\
           & 10-20\% & $65.4\pm5.2$ \\
           & 20-40\% & $60.7\pm5.2$ \\
           & 40-60\% & $54.1\pm5.2$ \\
           & 60-80\% & $44.6\pm5.9$ \\
\hline
      200 & 0-10\% & $28.4\pm5.5$ \\
          & 10-20\% & $27.7\pm5.1$ \\
          & 20-40\% & $27.4\pm4.9$ \\
          & 40-60\% & $22.9\pm4.9$ \\
          & 60-80\% & $18.2\pm4.5$ \\
\hline \hline
\end{tabular}
\end{table}

The Monte Carlo distribution generating method is used to obtain $p_{\rm T}$. Let $r_{\rm 1}$ denote the random numbers distributed uniformly in $[0,1]$. A series of values of $p_{\rm T}$ can be obtained by
\begin{equation}
\int_{0}^{p_{T}}f_{p_{T}}(p_{T})dp_{T}< r_{1}< \int_{0}^{p_{T}+dp_{T}}f_{p_{T}}(p_{T})dp_{T}.
\label{eq:2}
\end{equation}
Here, $f_{p_{T}}$ is the transverse momentum probability density function which is an alternative representation of the Tsallis distribution as follows:
\begin{equation}
f_{P_{T}}\left ( p_{T}\right )=\frac{1}{N}\frac{dN}{dp_{T}}=\int_{y_{min}}^{y_{max}}f\left ( p_{T},y\right )dy.
\label{eq:3}
\end{equation}
where $y_{max}$ and $y_{min}$ are the maximum and minimum rapidity, respectively.

Under the assumption of isotropic emission in the source rest frame, we use the Monte Carlo method to acquire the polar angle:
\begin{equation}
\theta=2{\arcsin\sqrt{r_{2}}}.
\label{eq:4}
\end{equation}
Here, $r_{\rm 2}$ denote the random numbers distributed uniformly in $[0,1]$. Thus, we can obtain a series of values of momentum and energy due to the momentum $p=\frac{p_{T}}{\sin\theta}$ and the energy $E={\sqrt{p^{2}+m_{\rm 0}^{2}}}$. Therefore, the corresponding values of rapidity can be obtained according to the definition of rapidity.

\vspace{1\baselineskip}

{\section{Results and discussion}}

{\subsection{Transverse momentum spectra}}

\begin{figure}
\setlength{\abovecaptionskip}{-1cm}
\includegraphics[angle=0,width=15.6cm]{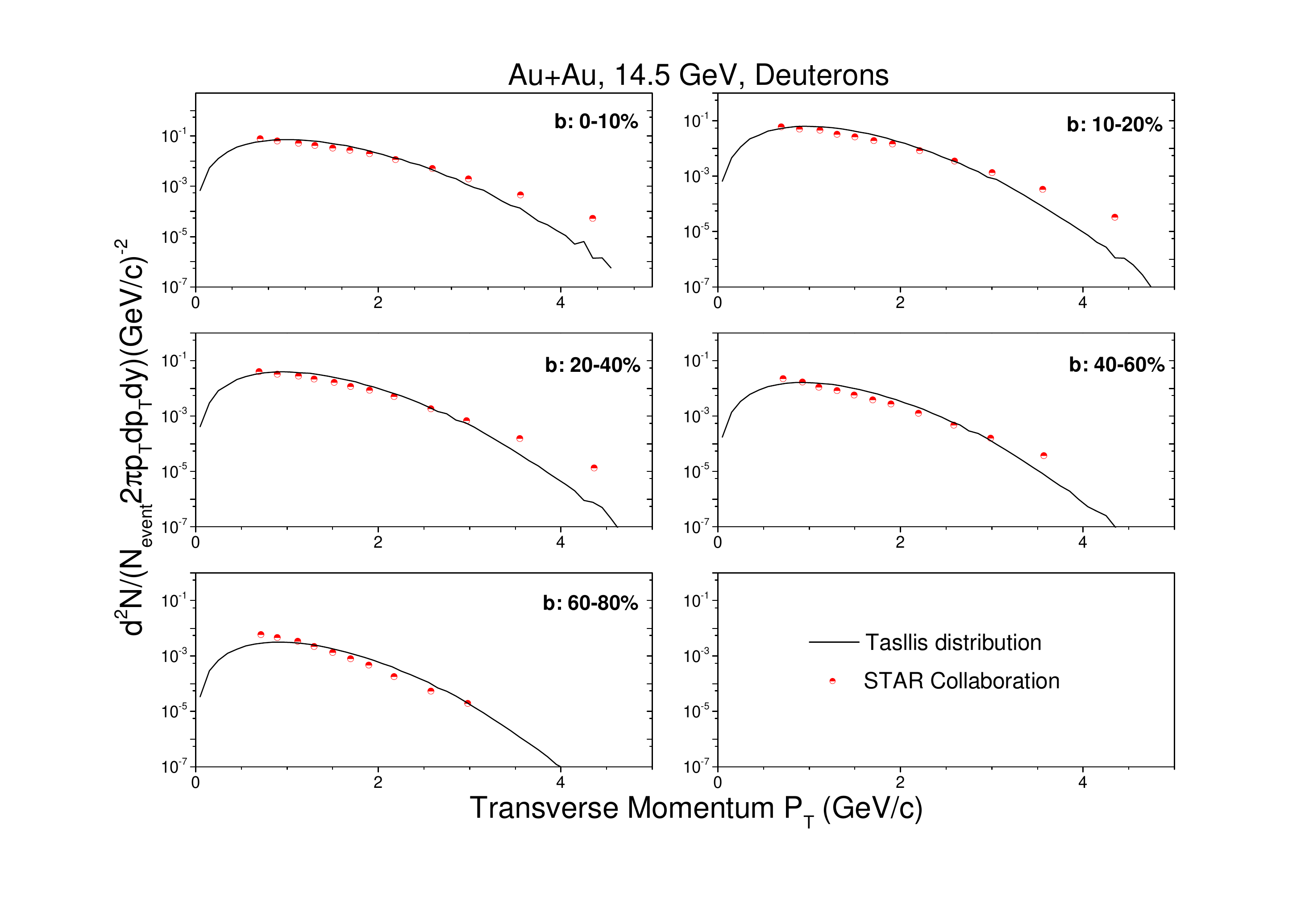}
\caption{Deuterons transverse momentum spectra in Au+Au collisions at $\sqrt{s_{NN}}$=14.5 GeV for $0-10\%$, $10-20\%$, $20-40\%$, $40-60\%$ and $60-80\%$ centralities. Calculations are shown by the solid lines. Experimental data taken from the STAR Collaboration \upcite{Adam20} are represented by the symbols.} \label{fig1}
\end{figure}

Fig.\ \ref{fig1} demonstrates mid-rapidity ($\vert{y}\vert$$<$0.3) transverse momentum spectra for deuterons in Au+Au collisions at $\sqrt{s_{NN}}$=14.5 GeV for $0-10\%$, $10-20\%$, $20-40\%$, $40-60\%$ and $60-80\%$ centralities. The symbols represent the experimental data of STAR Collaboration \upcite{Adam20}. The solid lines are our calculated results fitted by using the Tsallis distribution based on eq. (1) in the region of mid-rapidity. The values of the related parameters $T$ and $q$ are given in Table II along with the $\chi^{2}/dof$ ($\chi^{2}$ and number of degree of freedom). It is found that the calculations of the Tsallis distribution qualitatively describe the experimental data.

\begin{figure}
\setlength{\abovecaptionskip}{-1cm}
\includegraphics[angle=0,width=15.6cm]{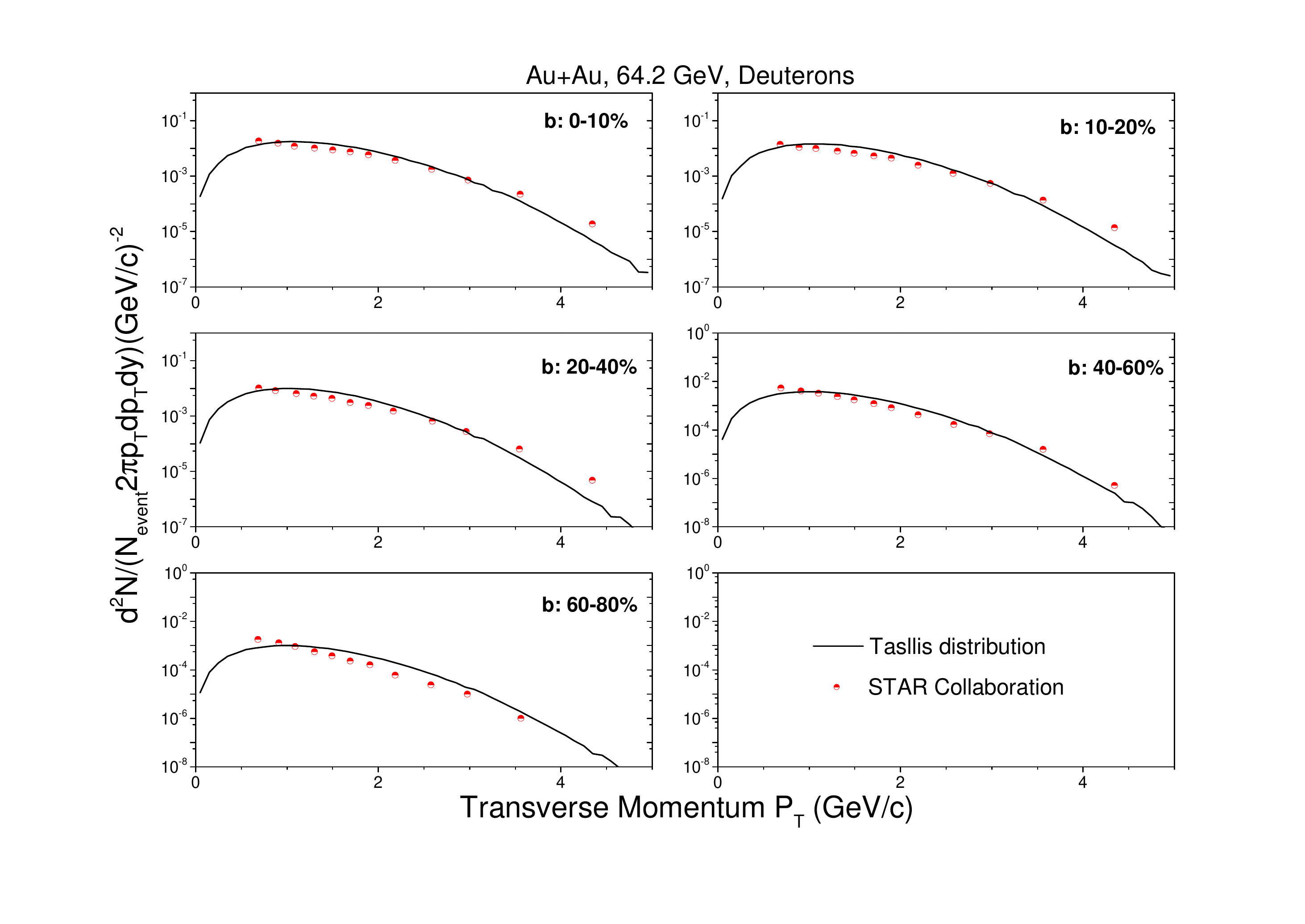}
\caption{Deuterons transverse momentum spectra in Au+Au collisions at $\sqrt{s_{NN}}$=62.4 GeV for $0-10\%$, $10-20\%$, $20-40\%$, $40-60\%$ and $60-80\%$ centralities. Calculations are shown by the solid lines. Experimental data taken from the STAR Collaboration \upcite{Adam20} are represented by the symbols.} \label{fig2}
\end{figure}

\begin{figure}
\setlength{\abovecaptionskip}{-1cm}
\includegraphics[angle=0,width=15.6cm]{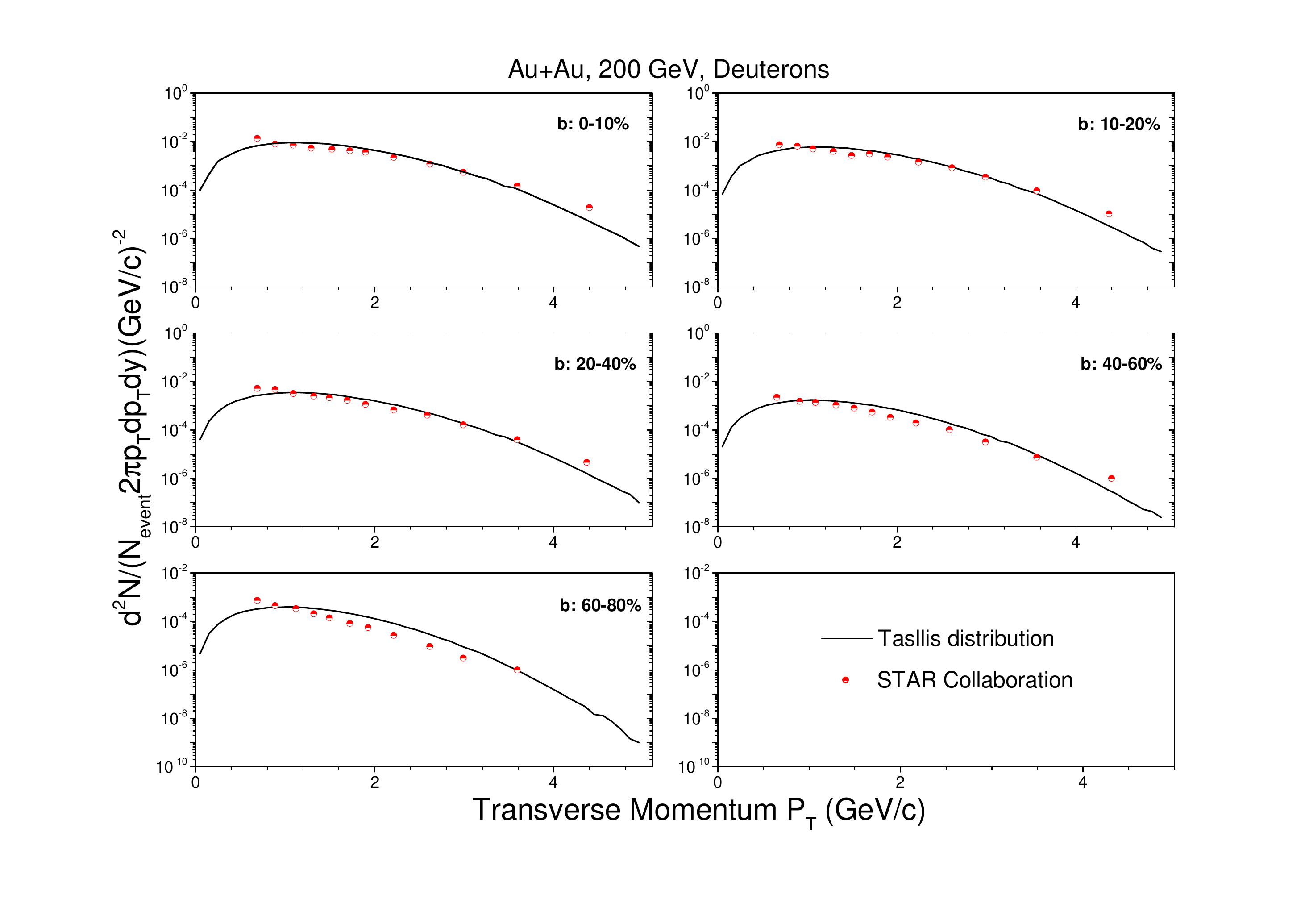}
\caption{Deuterons transverse momentum spectra in Au+Au collisions at $\sqrt{s_{NN}}$=200 GeV for $0-10\%$, $10-20\%$, $20-40\%$, $40-60\%$ and $60-80\%$ centralities. Calculations are shown by the solid lines. Experimental data taken from the STAR Collaboration \upcite{Adam20} are represented by the symbols.} \label{fig3}
\end{figure}

In Fig.\ \ref{fig2} and Fig.\ \ref{fig3}, the curves and symbols are similar to  Fig.\ \ref{fig1}. Figure\ \ref{fig2} demonstrates mid-rapidity ($\vert{y}\vert$$<$0.3) transverse momentum spectra for deuterons in Au+Au collisions at $\sqrt{s_{NN}}$=200 GeV for $0-10\%$, $10-20\%$, $20-40\%$, $40-60\%$ and $60-80\%$ centralities. The values of the related parameters $T$ and $q$ are given in Tables III and IV along with the $\chi^{2}/dof$. It is found that the calculations of the Tsallis distribution qualitatively describe the experimental data.

\begin{figure}
\setlength{\abovecaptionskip}{-1cm}
\includegraphics[angle=0,width=15.6cm]{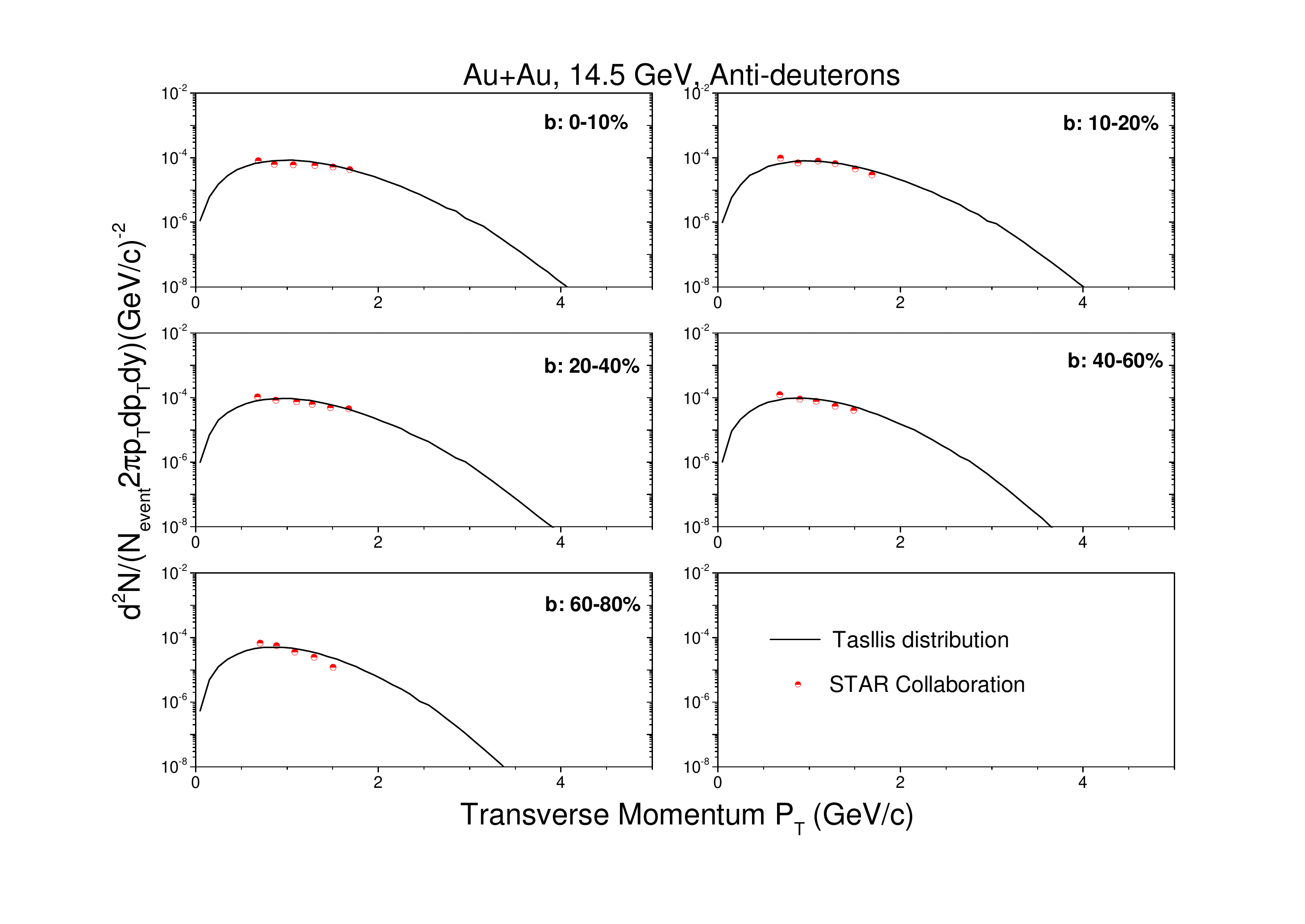}
\caption{Anti-deuterons transverse momentum spectra in Au+Au collisions at $\sqrt{s_{NN}}$=14.5 GeV for $0-10\%$, $10-20\%$, $20-40\%$, $40-60\%$ and $60-80\%$ centralities. Calculations are shown by the solid lines. Experimental data taken from the STAR Collaboration \upcite{Adam20} are represented by the symbols.} \label{fig4}
\end{figure}

\begin{figure}
\setlength{\abovecaptionskip}{-1cm}
\includegraphics[angle=0,width=15.6cm]{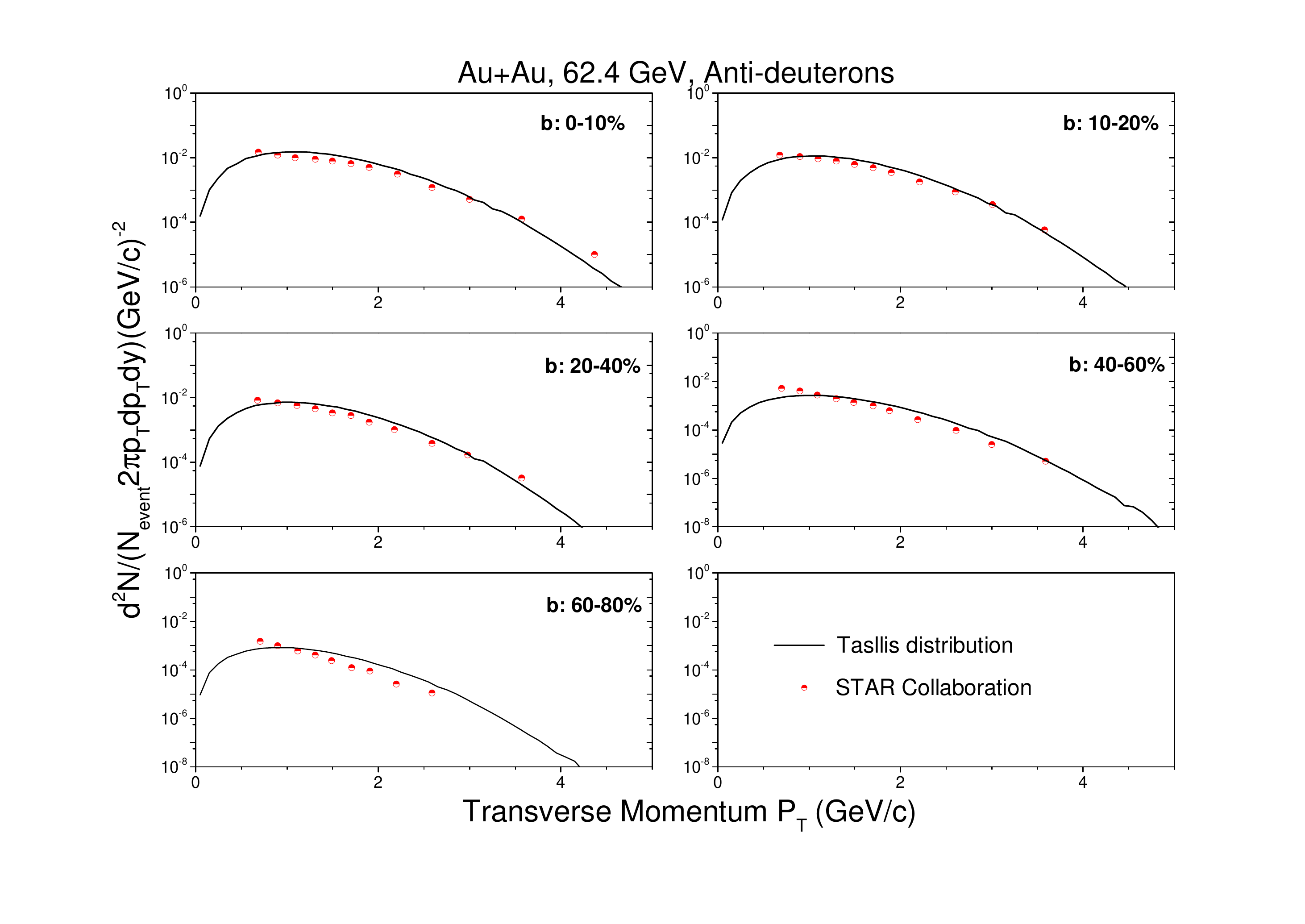}
\caption{Anti-deuterons transverse momentum spectra in Au+Au collisions at $\sqrt{s_{NN}}$=62.4 GeV for $0-10\%$, $10-20\%$, $20-40\%$, $40-60\%$ and $60-80\%$ centralities. Calculations are shown by the solid lines. Experimental data taken from the STAR Collaboration \upcite{Adam20} are represented by the symbols.} \label{fig5}
\end{figure}

\begin{figure}
\setlength{\abovecaptionskip}{-1cm}
\includegraphics[angle=0,width=15.6cm]{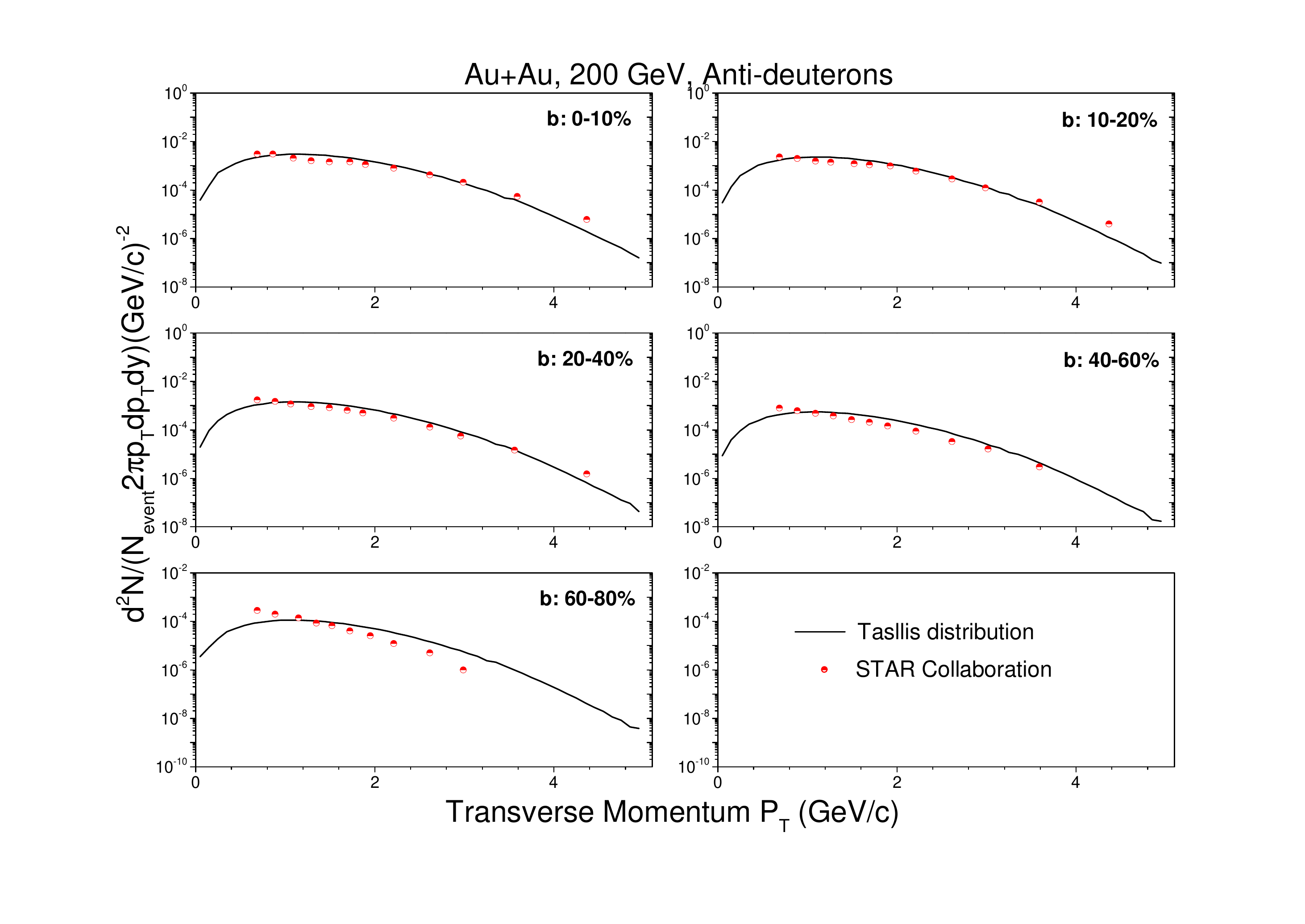}
\caption{Anti-deuterons transverse momentum spectra in Au+Au collisions at $\sqrt{s_{NN}}$=200 GeV for $0-10\%$, $10-20\%$, $20-40\%$, $40-60\%$ and $60-80\%$ centralities. Calculations are shown by the solid lines. Experimental data taken from the STAR Collaboration \upcite{Adam20} are represented by the symbols.} \label{fig6}
\end{figure}

Figure\ \ref{fig4}, Figure\ \ref{fig5} and Figure\ \ref{fig6} demonstrate the mid-rapidity ($\vert{y}\vert$$<$0.3) transverse momentum spectra for anti-deuterons in Au+Au collisions at $\sqrt{s_{NN}}$=14.5, 62.4 and 200 GeV for $0-10\%$, $10-20\%$, $20-40\%$, $40-60\%$ and $60-80\%$ centralities. The curves and symbols are similar to  Fig.\ \ref{fig1}. One can see that the calculations also can describe approximately the experimental data of anti-deuterons with different centrality intervals of event. The values of the related parameters $T$ and $q$ are given in Tables II, III, and IV.

\begin{table}
\caption{Values of $T$, $q$, and $\chi^{2}/dof$ corresponding to the curves in Au+Au collisions at $\sqrt{s_{NN}}$=14.5 GeV for $0-10\%$, $10-20\%$, $20-40\%$, $40-60\%$ and $60-80\%$ centralities. The 'Ratios' is the average ratios of experimental data to model.} \label{Table2}
\begin{tabular}{ p{1.5cm}<{\centering} p{1.5cm}<{\centering} p{3cm}<{\centering} p{3cm}<{\centering} p{3cm}<{\centering} p{2cm}<{\centering} p{2cm}<{\centering}}
\hline \hline
      Figure & Type 1 & Type 2 & T (GeV) & q & $\chi^{2}/dof$ & Ratios \\
\hline
      Fig. 1 & d & 0-10\% & $0.507\pm0.002$ & $1.125\pm0.017$ & 0.055 & 0.805 \\
             &   & 10-20\% & $0.487\pm0.011$ & $1.145\pm0.166$ & 0.053 & 0.742 \\
              &   & 20-40\% & $0.467\pm0.054$ & $1.165\pm0.084$ & 0.119 & 0.788 \\
              &   & 40-60\% & $0.427\pm0.008$ & $1.185\pm0.045$ & 0.150 & 0.848 \\
              &   & 60-80\% & $0.407\pm0.001$ & $1.205\pm0.007$ & 0.639 & 1.191 \\
\hline
      Fig. 4 & $\bar{d}$ & 0-10\% & $0.507\pm0.001$ & $1.105\pm0.001$ & 0.564 & 0.884 \\
              &   & 10-20\% & $0.487\pm0.001$ & $1.125\pm0.001$ & 0.239 & 1.005 \\
              &   & 20-40\% & $0.447\pm0.001$ & $1.145\pm0.001$ & 0.331 & 0.939 \\
              &   & 40-60\% & $0.387\pm0.001$ & $1.165\pm0.001$ & 0.619 & 0.914 \\
              &   & 60-80\% & $0.347\pm0.001$ & $1.185\pm0.001$ & 1.508 & 0.960 \\
\hline \hline
\end{tabular}
\end{table}

\begin{table}
\caption{Values of $T$, $q$, and $\chi^{2}/dof$ corresponding to the curves in Au+Au collisions at $\sqrt{s_{NN}}$=62.4 GeV for $0-10\%$, $10-20\%$, $20-40\%$, $40-60\%$ and $60-80\%$ centralities. The 'Ratios' is the average ratios of experimental data to model.} \label{Table3}
\begin{tabular}{ p{1.5cm}<{\centering} p{1.5cm}<{\centering} p{3cm}<{\centering} p{3cm}<{\centering} p{3cm}<{\centering} p{2cm}<{\centering} p{2cm}<{\centering}}
\hline \hline
      Figure & Type 1 & Type 2 & T (GeV) & q & $\chi^{2}/dof$ & Ratios \\
\hline
      Fig. 2 & d & 0-10\% & $0.607\pm0.008$ & $1.135\pm0.051$ & 0.037 & 0.785 \\
             &   & 10-20\% & $0.587\pm0.004$ & $1.155\pm0.038$ & 0.061 & 0.711 \\
              &   & 20-40\% & $0.527\pm0.006$ & $1.175\pm0.022$ & 0.124 & 0.731 \\
              &   & 40-60\% & $0.507\pm0.003$ & $1.195\pm0.010$ & 0.107 & 0.887 \\
              &   & 60-80\% & $0.487\pm0.001$ & $1.215\pm0.003$ & 0.274 & 0.910 \\
\hline
      Fig. 5 & $\bar{d}$ & 0-10\% & $0.607\pm0.001$ & $1.135\pm0.005$ & 2.527 & 0.712 \\
              &   & 10-20\% & $0.567\pm0.001$ & $1.155\pm0.005$ & 1.464 & 0.834 \\
              &   & 20-40\% & $0.527\pm0.001$ & $1.175\pm0.003$ & 2.231 & 0.833 \\
              &   & 40-60\% & $0.507\pm0.001$ & $1.195\pm0.003$ & 2.099 & 1.095 \\
              &   & 60-80\% & $0.407\pm0.002$ & $1.215\pm0.001$ & 3.303 & 0.966 \\
\hline \hline
\end{tabular}
\end{table}

\begin{table}
\caption{Values of $T$, $q$, and $\chi^{2}/dof$ corresponding to the curves in Au+Au collisions at $\sqrt{s_{NN}}$=200 GeV for $0-10\%$, $10-20\%$, $20-40\%$, $40-60\%$ and $60-80\%$ centralities.The 'Ratios' is the average ratios of experimental data to model.} \label{Table4}
\begin{tabular}{ p{1.5cm}<{\centering} p{1.5cm}<{\centering} p{3cm}<{\centering} p{3cm}<{\centering} p{3cm}<{\centering} p{2cm}<{\centering} p{2cm}<{\centering}}
\hline \hline
      Figure & Type 1 & Type 2 & T (GeV) & q & $\chi^{2}/dof$ & Ratios \\
\hline
      Fig. 3 & d & 0-10\% & $0.667\pm0.004$ & $1.145\pm0.021$ & 0.069 & 0.889 \\
             &   & 10-20\% & $0.647\pm0.004$ & $1.175\pm0.017$ & 0.040 & 0.847 \\
              &   & 20-40\% & $0.627\pm0.008$ & $1.195\pm0.036$ & 0.004 & 0.989 \\
              &   & 40-60\% & $0.567\pm0.001$ & $1.215\pm0.006$ & 0.048 & 0.782 \\
              &   & 60-80\% & $0.507\pm0.001$ & $1.235\pm0.003$ & 0.063 & 0.906 \\
\hline
      Fig. 6 & $\bar{d}$ & 0-10\% & $0.667\pm0.001$ & $1.145\pm0.005$ & 0.086 & 0.795 \\
              &   & 10-20\% & $0.647\pm0.001$ & $1.165\pm0.005$ & 0.047 & 0.770 \\
              &   & 20-40\% & $0.627\pm0.001$ & $1.195\pm0.004$ & 0.048 & 0.802 \\
              &   & 40-60\% & $0.607\pm0.001$ & $1.215\pm0.002$ & 0.055 & 0.853 \\
              &   & 60-80\% & $0.597\pm0.001$ & $1.255\pm0.001$ & 0.206 & 1.183 \\
\hline \hline
\end{tabular}
\end{table}

{\subsection{Average transverse momenta}}

\begin{figure}
\setlength{\abovecaptionskip}{-1cm}
\includegraphics[angle=0,width=15.6cm]{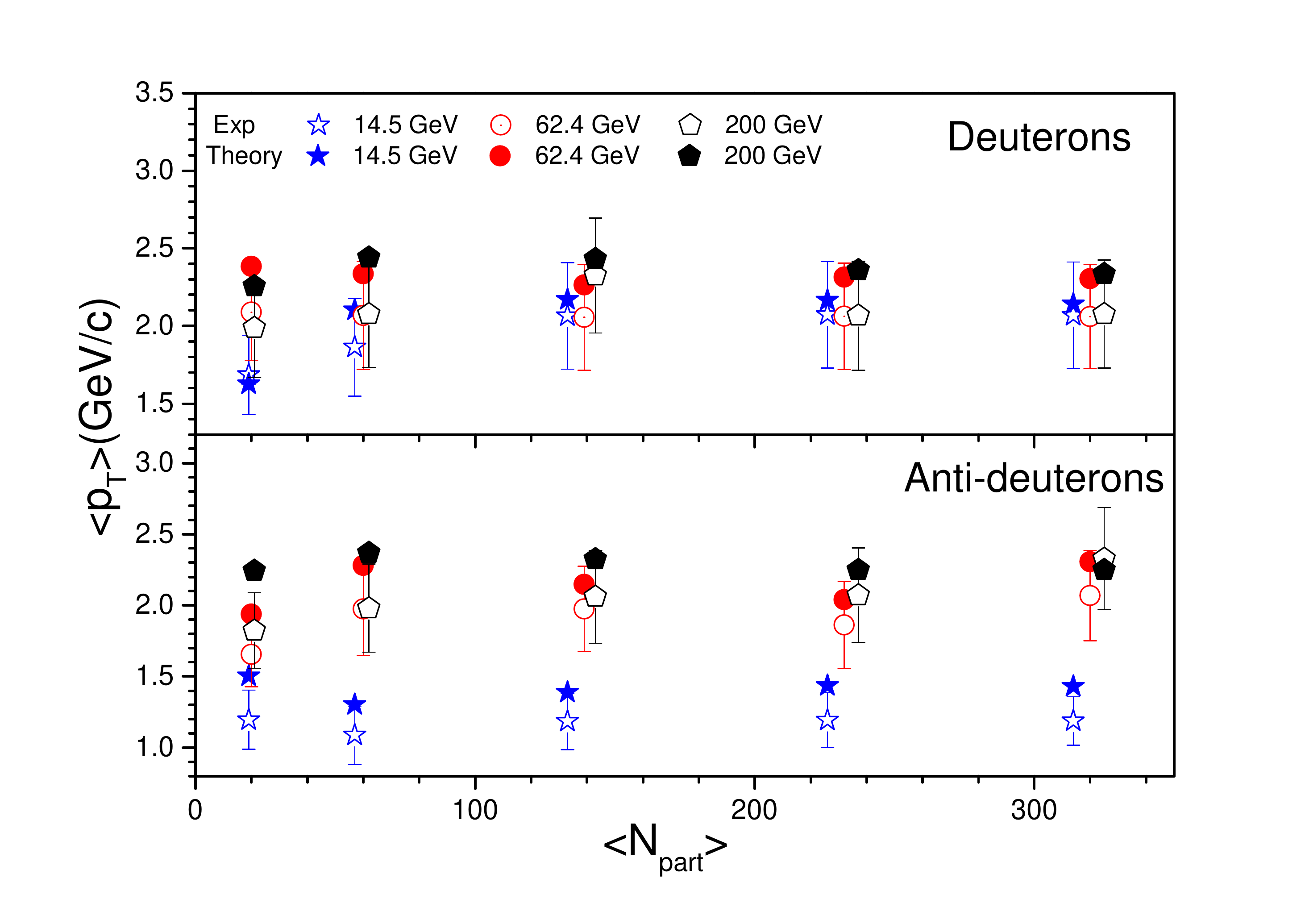}
\caption{Deuterons and anti-deuterons average transverse momenta ($\left\langle{p_{T}}\right\rangle$) as a function of $\left\langle{N_{part}}\right\rangle$ at mid-rapidity ($\vert{y}\vert$$<$0.3) for $\sqrt{s_{NN}}$=14.5, 62.4 and 200 GeV. Calculations are shown by the solid symbols. Experimental data taken from the Fig.\ \ref{fig1}-\ \ref{fig6} are represented by the hollow symbols.} \label{fig7}
\end{figure}

Figure\ \ref{fig7} presents the centrality dependence average transverse momenta ($\left\langle{p_{T}}\right\rangle$) of deuterons and anti-deuterons at the mid-rapidity ($\vert{y}\vert$$<$0.3) for $\sqrt{s_{NN}}$=14.5, 62.4 and 200 GeV. The hollow symbols are the experimental data, and the solid symbols are the calculations of the Tsallis distribution. The calculations can be obtained by
\begin{equation}
\left \langle {p_{T}}\right \rangle=\frac{\sum {p_{T1}}{\alpha }}{\sum {\alpha} }.
\label{eq:5}
\end{equation}
Here, ${p_{T1}}$ is the value of transverse momentum corresponding to the experimental data, and $\alpha$ is the value of
$\frac{d^{2}N}{{N_{event}}2\pi {p_{T}}d{p_{T}}dy}$ that corresponds to the ${p_{T1}}$.

In this figure, One observes that the calculations can describe the experimental data well in the range of the errors permitted. For deuterons, the values of average transverse momenta at different incident energy get closer with decrease of centrality percentage. It has indicated that the transverse excitation degree increases with collision centrality.

{\subsection{Dependence of parameters on number of participating nucleons}}

\begin{figure}
\setlength{\abovecaptionskip}{-0.5cm}
\includegraphics[angle=0,width=15.6cm]{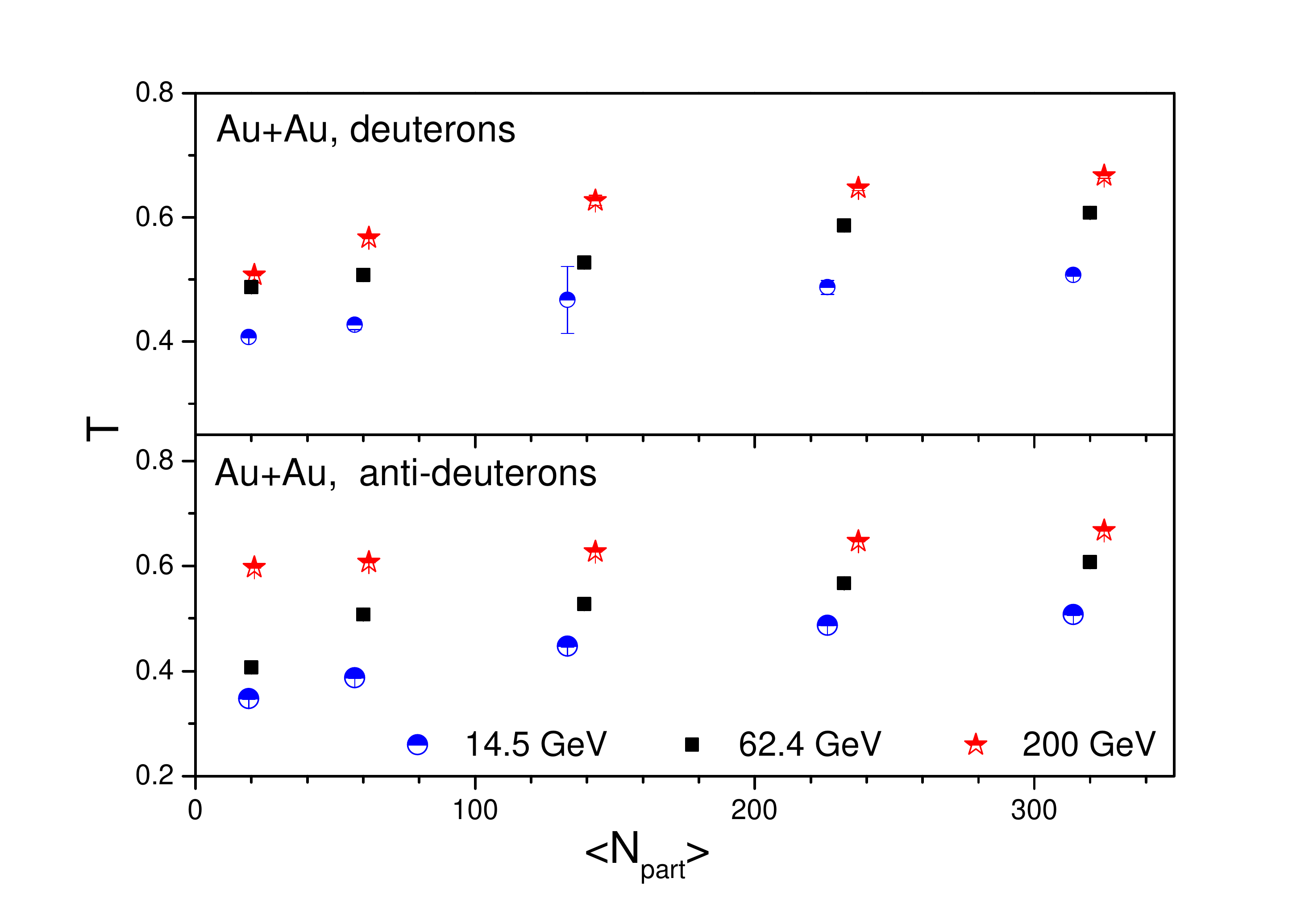}
\caption{Dependence of $T$ on the average number of participants for deuterons and anti-deuterons in events with different centrality intervals. The symbols represent the parameter values listed in Table 2, 3 and 4.} \label{fig8}
\end{figure}

\begin{figure}
\setlength{\abovecaptionskip}{-0.5cm}
\includegraphics[angle=0,width=15.6cm]{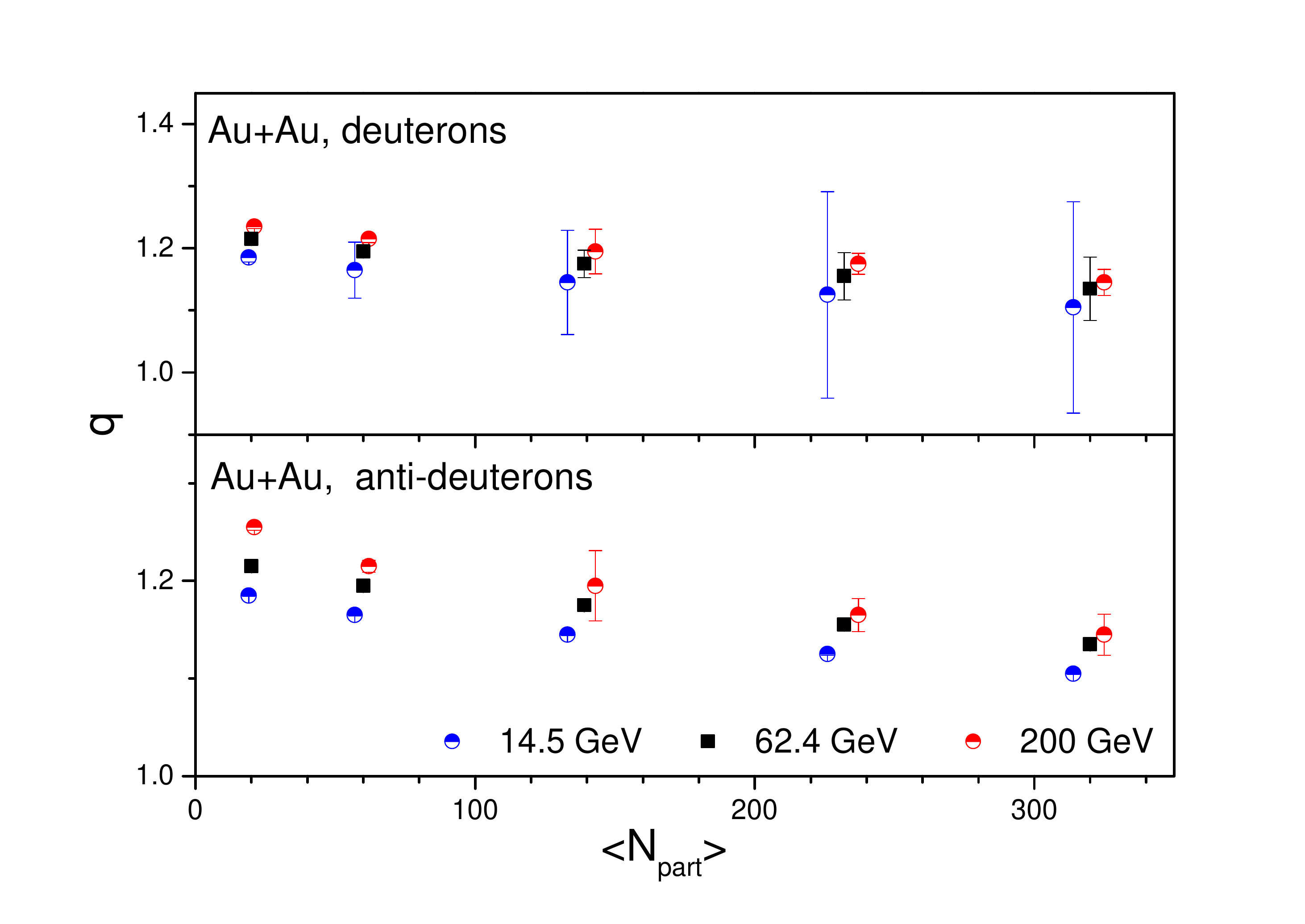}
\caption{Dependence of $q$ on the average number of participants for deuterons and anti-deuterons in events with different centrality intervals. The symbols represent the parameter values listed in Table 2, 3 and 4.} \label{fig9}
\end{figure}

Fig.\ \ref{fig8} and Fig.\ \ref{fig9} give the trends of parameters ($T$ and $q$) with the average number of participants for deuterons and anti-deuterons produced in Au+Au collision at the mid-rapidity ($\vert{y}\vert$$<$0.3) for $\sqrt{s_{NN}}$=14.5, 62.4 and 200 GeV. The symbols represent the parameter values extracted from Figs.\ \ref{fig1}-\ \ref{fig3} and Figs.\ \ref{fig4}-\ \ref{fig6} and listed in Tables II-IV.

From Fig.\ \ref{fig8} and Fig.\ \ref{fig9}, we can see that the values of $T$ parameters increase with decrease of centrality percentage, and the values of $q$ parameters increase with increase of centrality percentage. Entropy is a physical quantity that represents the degree of chaos in the system. When a central collision occurs, the motion law of the final state particles is complex, and the whole system is in a higher state of order, so the entropy value is small. In the central region where the collision occurs, with the increase of the intensity of the collision, the corresponding effective temperature increases. The dependence of effective temperature on collision energy increases with the increase of collision energy. Under the same collision parameters, the entropy increases with the increase of collision energy, indicating that the higher the collision energy is, the more different microscopic states the particle may have, and the more disordered the system becomes. The kinetic freeze-out temperature can be extracted from the effective temperature, the correlation between Kinetic freeze-out temperature and centrality will be focused in the future work.

\vspace{1\baselineskip}

{\section{Summary and Outlook}}

In summary, we have presented the transverse momentum distributions of (anti-)deuterons in Au+Au collisions at $\sqrt{s_{NN}}$=14.5, 62.4 and 200 GeV for $0-10\%$, $10-20\%$, $20-40\%$, $40-60\%$ and $60-80\%$ centralities. The Tsallis distribution in the multisource thermal model has been used in all calculations. Based on this model, we have investigated transverse momentum distributions of (anti-)deuterons, and the law about effective temperature and entropy with the centrality of collision. In conclusion, it can give the agreement between calculation results and the experimental data. The effective temperature extracted from $d$ and $\bar{d}$ increases with decrease of centrality percentage at the same incident energy, and  and the entropy index decreases with decrease of centrality percentage at the same incident energy. And at the same collision centrality, they increase with increase of incident energy. The effective temperature is not the actual temperature of the emission source. On the other hand the kinetic freeze-out temperature represents the temperature of the system when elastic collisions cease and hence is more realistic. We can extract the kinetic freezing temperature by the effective temperature. Then the kinetic freeze-out temperature and the evolution of time during the collision have yet to be studied in depth.

\vspace{1\baselineskip}

{\section*{Data Availability}}

The data used to support the findings of this study are included within the article and are cited at relevant places within the text as references.

{\section*{Conflict of Interests}}

The author declare that there is no conflict of interests regarding the publication of this paper.

{\section*{Acknowledgements}}

 This work was supported by the Introduction of Doctoral Starting Funds of Scientific Research of Guangxi University of Chinese Medicine under Grant No.2018BS024, the Natural Science Foundation of Guangxi Zhuangzu Autonomous Regions of China under Grant no. 2012GXNSFBA053011, Research support project of Guangxi institutions of higher learning No.200103YB071, and the Open Project of Guangxi Key Laboratory of Nuclear Physics and Nuclear Technology, No.NLK2020-03.

\end{document}